# Power in Transformation

## Creative Practices as Catalysts of Change

By Design and by Disaster

oekom science

# ARTIFICIAL INTELLIGENCE: BEYOND OCULARCENTRISM, THE NEW AGE OF HUMANS BEYOND THE SPECTACLE

## Mustapha El Moussaoui[1]

## Abstract

This paper explores the transformative impact of artificial intelligence (AI) on visual culture and its broader implications for contemporary society. The proliferation of machine learning models in generating visual content necessitates a critical reassessment of the relationship between reality and representation. AI-generated imagery not only challenges traditional conceptions of human creativity and perception but also intensifies the dominance of visual media in shaping public consciousness. By critiquing the reliance on vision as the primary mode of knowledge, this study examines how AI technologies blur the boundaries between reality and artificial constructs, deepening societal alienation.

To illustrate these dynamics, the paper presents an experiment conducted in Bolzano, Italy, where six distinct visual scenarios for an urban redevelopment project were created. Public engagement with these scenarios revealed a strong preference for visually striking AI-generated images, often at the expense of addressing real-life challenges, underscoring the influence of the "spectacle" in shaping perceptions and decisions.

The paper further investigates the role of AI in accelerating the commodification of images, perpetuating existing power structures, and raising critical questions about the human role in creating and interpreting visual media. Ultimately, this work calls for a re-evaluation of the societal implications of AI-driven visual culture, as it redefines the dynamics of observation, meaning, and agency.

1    Faculty of Design and Art, Free University of Bolzano  |  mustapha.elmoussaoui@unibz.it

# Introduction: The Rise of AI and the Crisis of Human Perception

The rapid ascent of artificial intelligence (AI) into the fabric of contemporary life marks a profound shift in how humanity perceives, interacts with, and constructs reality. In an era characterised by the proliferation of machine learning algorithms and neural networks, particularly those responsible for generating visual content, we are compelled to reconsider the very foundations of our epistemological and ontological frameworks. This paper seeks to interrogate the implications of AI's encroachment on the visual domain, focusing on how technologies such as Diffusions and Generative Adversarial Networks (GANs) are not merely tools but agents of transformation in our cultural and social landscapes.

At the heart of this analysis lies the critique of ocularcentrism, the privileging of vision over other senses in the Western tradition. The Western intellectual tradition has consistently elevated sight as the primary conduit to knowledge. Vision has historically been associated with objectivity, clarity, and truth, thereby establishing a hierarchical relationship between the observer and the observed (Jay, 1993; Levin, 1993).

However, this ocularcentric paradigm has also been complicit in the construction of power dynamics, where control over visual representation becomes a means of social domination.

The spectacle extends its critique into the realm of modern society, where the dominance of media representations shapes not only public perception but the very nature of reality itself. Here, images transform genuine social relations into abstracted, commodified forms. Individuals are distanced from authentic experiences as their engagement with the world becomes mediated through the lens of mass-produced visuals (Debord, 1967). This dynamic mirrors the broader mechanisms of social control, where vision operates as a tool for maintaining the existing power structures. The production and dissemination of images become acts of domination, serving the interests of those who hold cultural and economic power. Through the control of media and visual narratives, the ruling class subtly imposes its ideology, which the masses internalise as a natural view of the world. This process reflects the subtle workings of hegemony, where power is maintained not through overt coercion but through the normalisation of dominant perspectives, effectively shaping consciousness and limiting critical thought (Gramsci, 1971).

Moreover, the spectacle transforms vision into a means of reinforcing these hierarchies. Images, far from being neutral, encode values and beliefs that sustain the so-cial order, masking inequality and legitimising the status quo. The ideological work of media goes beyond representation, functioning as a site where consent to power is manufactured and sustained. The continuous circulation of these images integrates individuals into a system where their perceptions align with the interests of the powerful, deepening their alienation from both themselves and others (Hall, 1980).

Enter AI, a force that has begun to unravel the very fabric of these established paradigms. With the advent of GANs and technologies like Stable Diffusion, we are witnessing a dramatic acceleration in the production and consumption of images (Leach, 2022). These AI systems can generate visual content with a speed and aesthetic sophistication previously unimaginable, challenging the human monopoly on creative visual production. The implications of this shift are profound: If machines can produce images (videos, audios, etc.)—those central elements of the spectacle—faster and perhaps more convincingly than humans, what remains of our role in the creation and interpretation of visual culture?

This paper argues that AI does not merely replicate human visual capabilities but fundamentally reconfigures the relationship between the observer and the observed. The "hallucinations" produced by AI systems introduce a novel form of spectacle, one that blurs the boundaries between reality and fiction, and between human and machine-generated aesthetics. This compels us to reevaluate traditional approaches to visual representation and critically examine ocularcentric paradigms.

To illustrate this argument, the paper presents an experiment conducted in an urban setting, recreating six different visual scenarios. Five of these scenarios were generated through AI-based "hallucinations," while one was derived from a phenomenological study addressing real-life problems. These six scenarios were then exhibited to the public, inviting participants to select their preferred future public space. The results of this experiment are analysed to highlight the advantages and limitations of AI-generated visuals and their implications for ocularcentrism.

## Ocularcentrism: The Dominance of Vision in Western Thought

The primacy of vision in Western thought is not merely an incidental consequence of human biology; it is a deeply ingrained cultural construct that has shaped the epistemological and ontological frameworks of Western civilisation (Levin, 1993). Ocularcentrism, the privileging of sight over other senses, has long dictated how knowledge is produced, understood, and disseminated, often at the expense of a more holistic sensory experience (Pallasmaa, 1996).

Juhani Pallasmaa's, *The Eyes of the Skin: Architecture and the Senses* (1996), provides a crucial lens through which to critique this visual dominance. Pallasmaa argues that Western culture, particularly since the Enlightenment, has been dominated by an "ocularcentric" bias that marginalises the other senses, resulting in an impoverished experience of the world. He contends that this over-reliance on vision has led to an alienation of the body and a neglect of the full spectrum of human sensory experience, which is essential for a deeper, more embodied understanding of reality. In architecture, for instance, this visual bias manifests in designs that prioritise visual impact over tactile, auditory, and olfactory experiences, thereby reducing the richness of human interaction with space (El Moussaoui, 2020).

This critique of ocularcentrism is not new; its roots can be traced back to the inception of Western philosophy. Plato's allegory of the cave draws a clear distinction between those who live in the world of shadows—mere representations of reality—and those who manage to break free and see beyond these illusions. The spectacle, in this sense, traps individuals within a world of appearances, preventing them from accessing the truth behind these shadows. When one begins to see beyond the spectacle, they encounter reality as it truly is, untouched by the manufactured layers of deception that dominate their vision (Plato, Book VII, 2008).

This transformation of perception became even more pronounced with the introduction of religious imagery. Through sculptures and representations, the divine was inserted into the spatio-temporal realm, rendering the divine into a physical, visual form. This act of embodying the divine within a visual medium shifted the focus of apprehension from the spiritual to the material, emphasising the role of vision in understanding the world, although some scholars argue against reductionist models of "the gaze," representing that vision is not something that occurs in abstraction, but is a fundamental way of embodying the human self (Morgan, 2012).

Nevertheless, the rise of ocularcentrism has not gone unchallenged, as some have critiqued the Cartesian separation of mind and body, advocating for a more integrated understanding of perception that includes all the senses, emphasising the embodied nature of perception, arguing that vision is not a detached, objective process but one deeply intertwined with our bodily experiences and situated in the world (Merleau-Ponty, 2012; Heidegger, 1962).

In the contemporary era, the critique of ocularcentrism has been expanded by theorists who examine the intersection of technology and visual culture. Some philsophers explored how the advent of modern technologies, such as photography and cinema, reconfigured the relationship between the observer and the observed (Crary, 1990). Jonathan Crary argues that these technologies have not only transformed the way we see but have also altered the very nature of vision itself, making it increasingly mediated and commodified.

## The Spectacle: Social Control and the Mediated Reality

In the mid-20th century, Guy Debord's *The Society of the Spectacle* (1967) presented a radical critique of modern society, positing that in an era dominated by mass media and consumer capitalism, the spectacle has become the central organising principle of social life. The spectacle, according to Debord, is not merely a collection of images but a social relation mediated by images. It is a world where the representation of reality takes precedence over reality itself, leading to a profound alienation of individuals from authentic social relations. Debord's analysis begins with the assertion that "in societies where modern conditions of production prevail, all of life presents itself as an immense accumulation of spectacles." The spectacle, he argues, is a manifestation of the commodity fetishism described by Marx, where the image of a product becomes more important than its use-value. In the spectacle, reality is fragmented and mediated through images that serve the interests of the ruling class, perpetuating a passive consumption of reality rather than active engagement with it. The spectacle thus operates as a form of social control, reinforcing existing power structures by keeping individuals isolated and distracted.

The rise of digital technologies has only intensified the spectacle's reach and power. Today, AI-driven algorithms curate and generate vast amounts of visual content, shaping the way individuals perceive and interact with the world. The proliferation of platforms like Instagram, TikTok, and YouTube exemplifies this shift, where the spectacle is not only consumed but also actively produced by users who become both the creators and the subjects of mediated reality (Kollyri & Milioni, 2022; Armano & Briziarelli, 2017). Here, AI plays a crucial role in amplifying the spectacle, as algorithms determine what content is seen, liked, and shared, thereby reinforcing particular narratives and images over others.

One of the most significant developments in this regard is the advent of Generative Adversarial Networks (GANs) and other AI technologies that can create hyper-realistic images and videos. These technologies exemplify what Jean Baudrillard (1981) described as the "hyperreal"—a state where the distinction between reality and its representation becomes blurred, leading to a situation where simulations are more real than reality itself. Baudrillard's concept of the hyperreal is a natural extension of Debord's spectacle, wherein AI-generated content no longer merely represents reality but creates new realities that are consumed as authentic experiences.

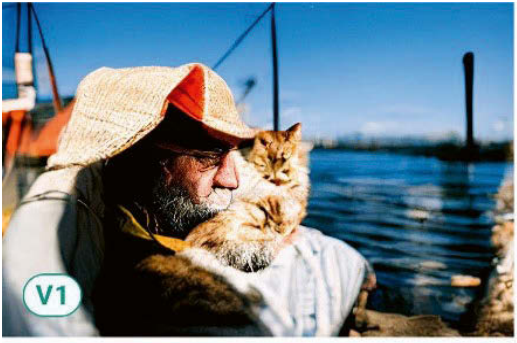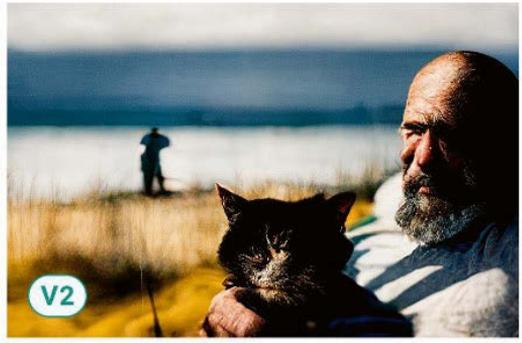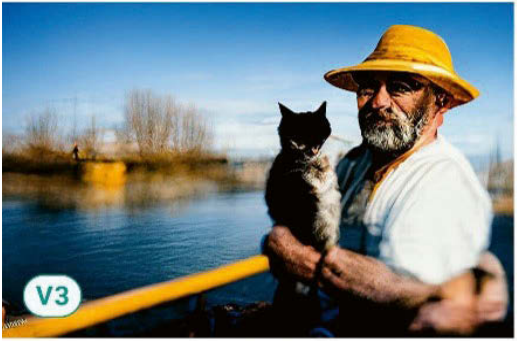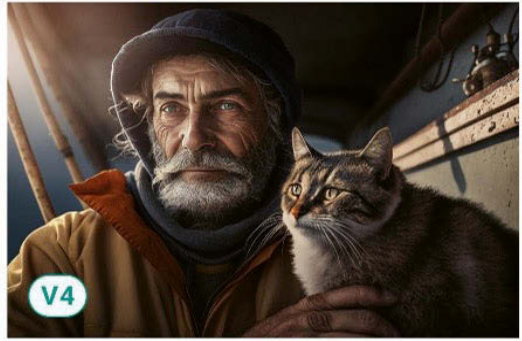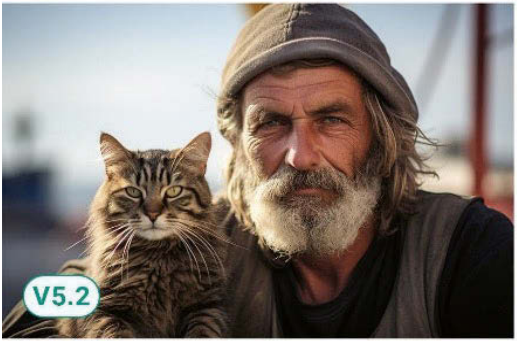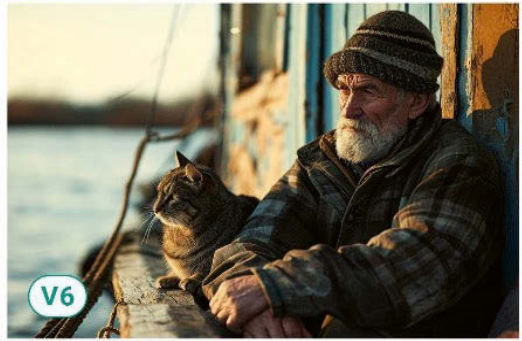

Figure 1 Midjourney outcome variable between February 2022 and December 2023 (Source: Donastre, 2024)

These AI-generated images and videos, which can be indistinguishable from real footage, complicate our understanding of truth and authenticity in the digital age. For instance, deepfakes—realistic AI-generated videos that can depict people saying or doing things they never actually said or did—represent a new frontier in the spectacle, where reality is not just mediated but entirely constructed by machines, and those who operate them. The implications for social control are profound, as these technologies can be used to manipulate public perception, spread misinformation, and reinforce hegemonic narratives with unprecedented efficiency and reach.

Contemporary theorists such as Shoshana Zuboff have explored how these developments are tied to what she terms "surveillance capitalism," a system where personal data is commodified and used to predict and influence behaviour (2019). Zuboff argues that the collection and analysis of vast amounts of data by tech companies enable a new form of social control, where individuals are subtly nudged towards certain behaviours through the manipulation of digital environments. This surveillance extends the logic of the spectacle, where reality is increasingly mediated through data-driven images and narratives designed to serve the interests of those who control the algorithms.

Moreover, the spectacle's reach into the digital domain has significant implications for the construction of identity and the experience of subjectivity. In a world where social media platforms encourage the constant curation of one's image, identity becomes a performance, a spectacle to be consumed by others, the "interpassivity", where individuals delegate their engagement with the world to external agents—whether they be images, technologies, or other people. In this context, the self becomes a spectacle, a mediated construction that obscures the complexities of real human experience (Pfaller, 2017).

## AI and the Deconstruction of Ocularcentrism through Image Generation, the Positive.

The advent of artificial intelligence, particularly in the realm of image generation, represents a profound challenge to the ocularcentric paradigm that has dominated Western thought for centuries. As discussed in earlier chapters, ocularcentrism—the privileging of vision as the primary mode of knowing—has long shaped our understanding of reality, knowledge, and power. However, AI technologies such as GANs and other deep learning models are now fundamentally altering this relationship between vision and knowledge.

These technologies do not merely replicate human visual capabilities; they transcend them, creating images that destabilise traditional notions of visuality, authenticity, and reality.

AI technologies, like GANs and Diffusions, are capable of producing images that are indistinguishable from photographs, yet these images do not originate from any physical reality. Instead, they are generated through complex algorithms that "hallucinate" visuals based on patterns learned from vast datasets. This process fundamentally redefines the traditional relationship between vision and reality. With rapid advancements in these technologies, they are now capable of generating highly realistic images accessible to the public. Within just two years a huge change took place (Figure 1), the realism of these generated images has grown exponentially.

One of the most striking aspects of AI-generated images is their ability to produce content that appears real but is entirely synthetic. This capacity for "deepfakes" and other forms of synthetic media calls into question the very nature of visual evidence. In the past, photographs and videos were often considered incontrovertible proof of events, serving as the basis for historical documentation, journalism, and legal evidence. However, as AI-generated images become more prevalent, the reliability of visual evidence is increasingly undermined. This shift has significant implications, as it forces us to reconsider the epistemological status of images and the role of vision in the construction of knowledge.

The deconstruction of ocularcentrism through AI image generation also has broader implications for the social and political dimensions of vision. As discussed in the previous chapter, the spectacle relies on the manipulation of visual images to maintain social control and reinforce dominant power structures. AI-generated images have the potential to both reinforce and subvert these structures. On one hand, the ability to create hyper-realistic images/videos,etc could be used to perpetuate the spectacle, creating new forms of visual manipulation that are even more insidious than those of the past. On the other hand, AI-generated images could also be used to challenge the spectacle, offering alternative visions that disrupt the dominant narratives imposed by those in power.

Furthermore, the deconstruction of ocularcentrism through AI-generated imagery opens up new possibilities for understanding the relationship between vision and the other senses. As Pallasmaa (1996) argues, the dominance of vision in Western culture has led to a neglect of the other senses, resulting in a disem-bodied experience of the world. AI technologies, by creating images that are not tied to any physical reality, encourage us to rethink the role of vision in our sensory hierarchy and consider how other senses can contribute to a more holistic understanding of the world. This shift has important implications for fields such as architecture, design, and media studies, where a more multisensory approach could lead to new forms of engagement with space, culture, and technology, as now more than ever, we truly understand that images' "representations" are widely used and can easily be manipulated, hence, pushing us to question the essence of the visual sense, the "representation," the why, to think beyond the ocularcentric.

## Beyond the Spectacle: AI, Rapid Image Creation, and the Transformation of Social Spaces

The integration of artificial intelligence into the fabric of visual culture marks a significant turning point in the evolution of social spaces, with a possibility to transcend the spectacle that has long dominated our mediated realities. At the core of this transformation is the ability of AI technologies, to produce images with unprecedented speed and sophistication. These technologies do not simply replicate existing visual forms; they generate new, often unexpected, aesthetic experiences that challenge the established norms of visual culture. The sheer velocity at which AI can create and disseminate images disrupts the temporal rhythm of the spectacle, which has traditionally relied on the gradual accumulation of images to construct and reinforce social realities. In this new paradigm, the rapidity of image production destabilises the spectacle's hold on the collective consciousness, opening up possibilities for the emergence of new social spaces that are less constrained by the dictates of commodified visuality.

One example of this phenomenon is the rise of virtual and augmented reality environments, where AI-generated images create immersive spaces that blur the boundaries between the real and the virtual in seconds. These new AI- Augmented reality environments challenge traditional notions of space by offering experiences that are both hyper-real and hyper-mediated, created in mere seconds to commodify the needs of the spectator. These new media environments reconfigure the relationship between the body and space, creating what he calls "mixed reality" spaces where the physical and digital coexist in a continuous feedback loop. In such spaces, the spectacle is not merely a passive consumption of images but an active engagement with a constantly evolving visual field, driven by AI's capacity to generate and transform images in real time.

## Scenarios and the Spectacle

In October 2023, an experiment was conducted in the Don Bosco region of Bolzano, Italy, to explore the implications of AI-generated imagery on urban development and community engagement. Six distinct scenarios were created, five of which were generated using AI tools, specifically MidJourney V6 and ComfyUI, employing GANs.

The first four AI-generated scenarios were inspired by a research project aligned with the 2030 Agenda for Sustainable Development and the European Green Deal. This project proposed four plausible trajectories for the region's future by 2030, each embodying distinct visions of sustainability and societal change. These scenarios included:

1. **Regional Consciousness:** "Our strength lies in tradition."

2. **Neo-Cosmopolitanism:** "Think global, act local."

3. **Individual Freedom:** "I am the architect of my own happiness."

4. **Green Innovation:** "There is a (technological) solution to everything." (Habicher et al., 2022)

5. **Dystopic Scenario**

6. **Real-Life Problematics**

GANs were employed to visually recreate these hypothetical futures, offering a compelling, image-driven representation of public spaces in the neighbourhood. These AI-generated visuals were juxtaposed with the author's phenomenological scenario, which addressed real-life urban challenges, including accessibility, safety, and social cohesion.

To ground the experiment in public participation, a community event was organised in Bolzano, where residents engaged with the six scenarios. This participatory approach aimed to involve the local population in reimagining their urban environment and to gauge their understanding of AI's role in shaping public spaces. The methodology combined qualitative and quantitative feedback tools to ensure diverse and inclusive participation, and the participants were informed about the concept, problems, and solutions proposed (Figure 2).

The event attracted 87 participants across a wide age range, including children, young adults, and the elderly, ensuring a broad spectrum of perspectives. After the scenarios and their contextual backgrounds were explained, participants were asked to select their preferred vision for the neighbourhood's redevelopment. Notably, 70 out of 87 participants (around 80%) gravitated toward the AI-generated scenarios that offered visually striking representations, often described as "spectacular." This preference highlighted the dominance of visual allure over pragmatic, problem-oriented solutions—reinforcing the argument that the "spectacle" exerts a powerful influence on decision-making (Figure 3).

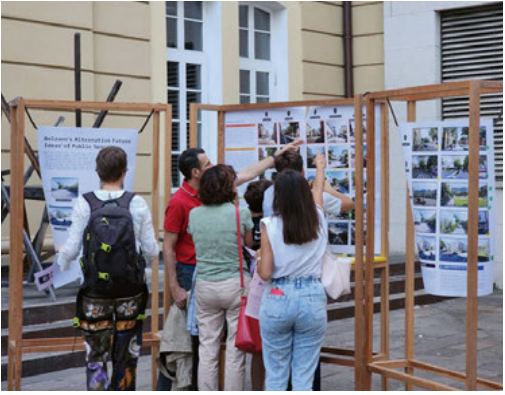

Figure 2 Participants making scenario choices.
Source : 2023 Author

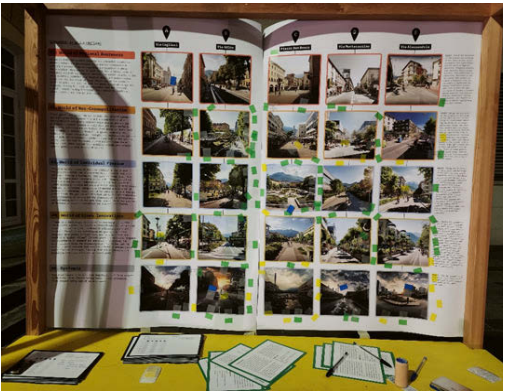

Figure 3 Some of the votes chosen by participants.
Source : 2023 Author

This outcome underscores a critical tension between visual appeal and substantive engagement with real urban issues. The experiment serves as a case study to interrogate how AI-generated imagery, while innovative, risks perpetuating ocularcentrism by prioritising aesthetics over lived realities in urban design. It also emphasises the need for critical frameworks to navigate the interplay between visual representation, community participation, and meaningful urban transformation.

# Human–Beings Where To?: The Future of Identity and Essence in the AI Age

The question of identity has always been central to philosophy, rooted in the quest to understand the self and its relation to the world. The advent of AI marks a critical juncture in the evolution of human identity and essence. As AI technologies become more deeply embedded in various aspects of daily life, they challenge fundamental understandings of identity, particularly when identity is considered a fluid social construct shaped by interaction with external forces, including technological advancements (Taylor, 1989; Giddens, 1991). The growing presence of AI raises profound questions about the nature of human identity, agency, and perception in the digital age.

The increasing convergence of humans and machines necessitates a rethinking of subjectivity. Traditional notions of the human as a bounded, autonomous entity are being supplanted by more fluid and distributed forms of subjectivity, incorporating both biological and technological elements. This posthuman condition challenges fixed definitions of identity, reframing it as constituted through networks of interactions that span organic and synthetic domains (Hayles, 1999).

The fluidity of identity was underscored in the experiment conducted in the Don Bo-

Mustapha El Moussaoui                                        45

sco region of Bolzano, Italy. By presenting six distinct scenarios—five generated using AI tools such as MidJourney V6 and ComfyUI, and one rooted in a phenomenological interpretation of real-life neighbourhood challenges. Public engagement with these scenarios revealed the dominance of the visual spectacle: 80% of the participants favoured AI-generated scenarios for their aesthetic appeal, even when these failed to address the actual problematics of the area, and they were hallucinative in their core. This preference highlighted the influence of AI-generated imagery on collective decision-making, reinforcing the argument that the "spectacle" continues to mediate identity and agency, often prioritising allure over substance.

These findings connect within broader discourses on AI and identity. Algorithmically driven experiences increasingly mediate human interactions, influencing perceptions and shaping how individuals construct their sense of self. Personalisation technologies, while designed to enhance user experiences, often reinforce biases and create echo chambers that constrain identity formation (Tufekci, 2017). Similarly, the allure of AI-generated imagery amplifies ocularcentrism, perpetuating the privileging of vision as the primary mode of understanding while sidelining deeper, multidimensional engagement with reality.

Yet, the future of identity and essence in the AI age is not predetermined. While AI presents significant challenges, it also offers opportunities for reimagining what it means to be human. The experiment in Bolzano serves as a reminder of the need to critically engage with AI's visual allure, ensuring that its potential is controlled for meaningful transformation rather than perpetuating new forms of control. The posthuman condition offers an opportunity to move beyond anthropocentric and dualistic frameworks, embracing instead a dynamic understanding of identity that incorporates both human and non-human actors (Braidotti, 2013).

In this context, the essence of humanity no longer resides in fixed characteristics but in the capacity for adaptation, creativity, and resilience amid change. AI, by challenging the dominance of the visual forces a reconsideration of what it means to be human. It calls for new epistemologies that transcend the limitations of the empirical, pushing us beyond the constructed narratives of power that have shaped our understanding, additionally to question and critically approach any narrative presented to us. This reimagining demands a holistic, multidimensional approach to knowledge, engaging all senses and perspectives to arrive at deeper truths, rather than submitting to constructed narratives imposed on us. The experiment in Bolzano demonstrates both the potential and the peril of AI in reshaping identity and perception. As AI creates representa-tions beyond human imagination, we are called to question what is truly necessary and existential. To avoid being drawn into a new, more deterministic spectacle, we must critically engage with this shift, ensuring it becomes an opportunity for genuine human evolution rather than another mechanism of control.

# Bibliography

Armano, E., & Briziarelli, M. (2017). The Spectacle 2.0: Reading Debord in the Context of Digital Capitalism. University of Westminster Press.

Baudrillard, J. (1994). Simulacra and Simulation (S. F. Glaser, Trans.). University of Michigan Press. (Original work published 1981)

Braidotti, R. (2013). The Posthuman. Polity Press.

Crary, J. (1990). Techniques of the Observer: On Vision and Modernity in the Nineteenth Century. MIT Press.

Debord, G. (1994). The Society of the Spectacle (D. Nicholson-Smith, Trans.). Zone Books. (Original work published 1967)

El Moussaoui, M. (2020). The ocular-centric obsession of contemporary societies. Civil Engineering and Architecture, 8(6), 566–577. https://doi.org/10.13189/cea.2020.080613

Giddens, A. (1991). Modernity and Self-Identity: Self and Society in the Late Modern Age. Stanford University Press.

Gramsci, A. (1971). Selections from the Prison Notebooks. International Publishers.

Hall, S. (1980). Encoding/decoding. In S. Hall, D. Hobson, A. Lowe, & P. Willis (Eds.), Culture, Media, Language (pp. 128–138). Hutchinson.

Hayles, N. K. (1999). How We Became Posthuman: Virtual Bodies in Cybernetics, Literature, and Informatics. University of Chicago Press.

Heidegger, M. (1962). Being and Time (J. Macquarrie & E. Robinson, Trans.). Harper & Row. (Original work published 1927)

Jay, M. (1993). Downcast Eyes: The Denigration of Vision in Twentieth-Century French Thought. University of California Press.

Kollyri, L., & Milioni, D. L. (2022). "Spectacular" user subjectivities on Instagram: A discursive interface analysis. Convergence: The International Journal of Research into New Media Technologies, 28(5), 1255–1273. https://doi.org/10.1177/13548565211050325

Leach, N. (2022). Architecture in the Age of Artificial Intelligence: An Introduction to AI for Architects. Bloomsbury Visual Arts.

Levin, D. M. (1993). Modernity and the Hegemony of Vision. University of California Press.

Merleau-Ponty, M. (2012). Phenomenology of Perception (D. A. Landes, Trans.). Routledge. (Original work published 1945)

Morgan, D. (2012). The Embodied Eye: Religious Visual Culture and the Social Life of Feeling. University of California Press.

Pallasmaa, J. (1996). The Eyes of the Skin: Architecture and the Senses. John Wiley & Sons.

Pfaller, R. (2017). Interpassivity: The Aesthetics of Delegated Enjoyment. Edinburgh University Press.

Plato. (2008). The Republic (R. Waterfield, Trans.). Oxford University Press. (Original work published ca. 380 BCE)

Taylor, C. (1989). Sources of the Self: The Making of the Modern Identity. Harvard University Press.

Tufekci, Z. (2017). Twitter and Tear Gas: The Power and Fragility of Networked Protest. Yale University Press.

Zuboff, S. (2019). The Age of Surveillance Capitalism: The Fight for a Human Future at the New Frontier of Power. PublicAffairs.

Habicher, D., Windegger, F., von der Gracht, H. A., & Pechlaner, H. (2022). Beyond the COVID-19 crisis: A research note on post-pandemic scenarios for South Tyrol 2030+. Technological Forecasting and Social Change, 180, 121749. https://doi.org/10.1016/j.techfore.2022.121749

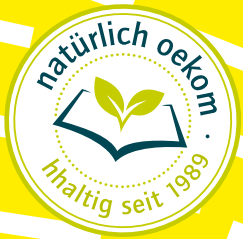